\documentclass[aps, prd, superscriptaddress, preprintnumbers, showpacs]{revtex4}
\usepackage[english]{babel}
\usepackage{graphicx,amsfonts,amsmath,amssymb,amstext}
\usepackage{float,wrapfig}
\usepackage{subfigure, psfrag}
\usepackage{dsfont}
\usepackage{epsfig}
\usepackage{color}
\usepackage{bm}
\usepackage{multirow,textcomp}
\usepackage{natbib}
\usepackage[normalem]{ulem}

\newcommand{\be}{\begin{equation}}
\newcommand{\ee}{\end{equation}}
\newcommand{\ba}{\begin{eqnarray}}
\newcommand{\ea}{\end{eqnarray}}

\definecolor{purple}{rgb}{0.8,0,0.6}

\makeatletter
\newcommand{\vast}{\bBigg@{2}}
\newcommand{\Vast}{\bBigg@{3}}
\makeatother

\begin{document}

\title{Magnetogenesis in Higgs inflation }
\date{\today}

\author{Mehran  Kamarpour}
\affiliation{Physics Faculty, Taras Shevchenko National University of Kyiv, 64/13, Volodymyrska str., 01601 Kyiv, Ukraine}

\begin{abstract}
We study the generation of magnetic fields in the Higgs inflation model with the axial coupling in order to break the conformal invariance of the Maxwell action and produce strong magnetic fields. We consider radiatively corrected Higgs inflation potential. In comparison to the Starobinsky potential, we obtain an extra term as a one loop correction and determine the
spectrum of generalized electromagnetic fields.
For two values of coupling parameter $\chi_{1}=5\times 10^{9}$ and $\chi_{1}=7.5\times 10^{9}$, the back-reaction is weak and our analysis is self-consistent.

\end{abstract}


\maketitle


\tableofcontents

\section{Introduction}
\label{sec-intro}

Recently magnetic fields were detected in the cosmic voids through the gamma-ray observations of distant blazars
\cite{Neronov:2010,Tavecchio:2010,Taylor:2011,Caprini:2015} with very large coherence scale $\lambda_{B}\gtrsim 1\,$Mpc. The origin of
these fields is a very intriguing problem which may shed light on the physical processes in the early Universe
\cite{Kronberg:1994,Grasso:2001,Widrow:2002,Giovannini:2004,Kandus:2011,Durrer:2013,Subramanian:2016}.
Combining these observations with the cosmic microwave background (CMB) data \cite{Planck:2015-pmf,Sutton:2017,Jedamzik:2018} constrains the 
strength of these magnetic fields to $10^{-17}\lesssim B_{0}\lesssim 10^{-9}\,$G.
The extremely large correlation length of magnetic fields observed in the cosmic voids strongly suggests that they were produced during
an inflationary stage of the evolution of the Universe because the inflationary magnetogenesis
Refs.~\cite{Turner:1988,Ratra:1992} can easily attain very large coherence length.

Since Maxwell's action is conformally invariant, the fluctuations of the electromagnetic field are not enhanced in the conformally flat 
inflationary background \cite{Parker:1968}. A standard way to break the conformal invariance is to introduce the interaction with scalar 
field or curvature scalar \cite{Turner:1988,Ratra:1992,Garretson:1992,Dolgov:1993}. One of them is the kinetic coupling of the 
electromagnetic field to the scalar inflation field via the term $f^{2}(\phi)F_{\mu\nu}F^{\mu\nu}$, which was proposed by Ratra \cite{Ratra:1992} and 
then studied in detail for different types of coupling functions in the literature
\cite{Giovannini:2001,Bamba:2004,Martin:2008,Demozzi:2009,Kanno:2009,Ferreira:2013,Ferreira:2014,Vilchinskii:2017}.

One of the essential features of the kinetic coupling model is the modification of the electromagnetic coupling constant. Indeed, since the 
standard electromagnetic Lagrangian is multiplied by $f^2$, one can rescale \cite{Demozzi:2009} the electromagnetic potential and absorb $f$.
As a result, the electric charges of particles effectively will depend on $f^{-1}$. Obviously, for small $f$, this leads to a strong 
coupling problem. Therefore, one should require $f\ge 1$ in order to avoid this problem during inflation. Further, since the 
inflation field and the scale factor change monotonously during inflation, it is natural to assume that the coupling function is a decreasing 
function during inflation which attains large values in the beginning. For a decreasing coupling function, it is also well known that the 
electric energy density dominates the magnetic one \cite{Martin:2008,Demozzi:2009,Vilchinskii:2017} and the energy density of the generated 
electric field may exceed that of the inflation field leading to the back-reaction problem.
In our paper, we consider the axial coupling $RF_{\mu\nu}\bar{F}^{\mu\nu}$ which violates parity and $\bar{F}^{\mu\nu}$ is the dual of $F^{\mu\nu}$. This term is presented in the Jordan frame. In the 
Einstein frame such term naturally produce non-trivial coupling between inflation and the electromagnetic field as we discuss in Sec. \ref{sec3}.
 
Inflation is a wonderful paradigmatic idea which naturally solves some very difficult problems of the hot Big Bang model 
\cite{Guth,Linde,Mukhanov}. Yet the nature of the inflation field is an open question. Usually it is assumed to be a scalar or pseudo-scalar 
field. The Standard Model of elementary particles contains only one fundamental scalar field. This is the Higgs boson. Therefore, it is a 
natural minimalist idea to try to employ the Higgs boson as the inflation field. Remarkably, this turned out to be a viable scenario if the 
Higgs field couples nonminimally to the curvature scalar $\xi h^2R/2$ with sufficiently strong coupling constant $\xi$. It was shown in
Ref.~\cite{Shaposhnikov:2008} that the non-minimal coupling $\xi$ ensures the flatness of the scalar potential in the Einstein frame at large
values of the Higgs field. The successful inflation consistent with the amplitude of the scalar perturbations in the CMB takes place for very 
large values $\xi$ of order $10^4$. This simplest and most economical scenario provides the graceful exit from inflation and predicts the tilt 
of the spectrum of the scalar perturbations $n_s \simeq 0.97$ and a very small tensor-to-scalar ratio $r \simeq 0.003$.
After inflationary period, the oscillations of the Higgs field about the Standard Model vacuum reheat the Universe producing the hot Big Bang
with temperature $10^{13-14}\,\mbox{GeV}$ \cite{Shaposhnikov:2009,Rubio:2009}.

Certainly, the quantum radiative corrections can significantly modify the form of the effective potential. This issue was carefully studied in 
the literature \cite{Barvinsky:2008,Magnin:2009,Wilczek:2009,Bezrukov:2009,Barvinsky:2009} and it was found that the experimentally observed 
mass of the Higgs boson is less than the critical value. Interestingly, however, it was shown also in a recent paper \cite{Shaposhnikov:2015} 
that the successful Higgs inflation can take place even if the Standard Model vacuum is metastable.

This paper is organized as follows. In Sec. \ref{sec-model} we consider the Higgs model. In Sec. \ref{sec3} we introduce axial coupling and finally in Sec. \ref{sec4} the relations for determining power spectra are considered. The summary of the obtained results is given in Sec.~\ref{sec-concl}.

\section{Radiatively corrected Higgs inflation}
\label{sec-model}

The Lagrangian in the Higgs inflation model is given by
\begin{equation}
\label{E1}
L=\sqrt{-g}\left[-f\left(h\right)R+\frac{1}{2}\left(\partial{h}\right)^{2}-U\left(h\right)-\Lambda\right],
\end{equation}
where $ U\left(h\right)=\frac{\lambda}{4}\left(h^{2}-v^{2}\right)^{2}$ is the Higgs potential of the Standard Model in unitary gauge, $2H^{\dagger}H=h^{2}$, and $ \Lambda $ is cosmological constant. Further, $f\left(h\right)=\frac{M^{2}_{p}+\xi_{h}h^{2}}{2} $. Here $M_{p} $ is the reduced Planck mass and $\xi_{h}$ is the non-minimal coupling constant.
Note that the Lagrangian is written in the Jordan frame. According to Ref. \cite{Shaposhnikov:2008}, we can avoid $\Lambda$ because its effect is negligible. We will work in the Einstein frame. For this, we perform the conformal transformation \cite{Subramanian:2016,Maeda:1989}
\begin{equation}
\label{E2}
\tilde{g}_{\mu\nu}=\Omega^{2}g_{\mu\nu}.
\end{equation}
This implies $\sqrt{-\tilde{g}}=\Omega^{4}\sqrt{-g} $ and $\tilde{g}^{\mu\nu}=\Omega^{-2}g^{\mu\nu}$. In order to rewrite Eq. (\ref{E1}) in the Einstein frame, we have the following relations: \cite{Shaposhnikov:2008,Maeda:1989}
\begin{equation}
\label{E3}
\frac{f\left(h\right)}{\Omega^{2}}=\frac{M^{2}_{p}}{2}.
\end{equation}
This implies the following relation between the conformal transformation and Higgs field: \cite{Shaposhnikov:2008,Maeda:1989}
\begin{equation}
\label{E4}
\Omega^{2}\left(h\right)=1+\frac{\xi_{h}h^{2}}{M^{2}_{p}}
\end{equation}
As a result, the Lagrangian density in the Einstein frame takes the form
\begin{equation}
\label{E5}
\tilde{L}=\sqrt{-\tilde{g}}\left[-\frac{M^{2}_{p}}{2}\tilde{R}+\frac{1}{2}\left(\frac{\Omega^{2}+\frac{6\xi^{2}_{h}h^{2}}{M^{2}_{p}}}{\Omega^{4}}\right)\tilde{g}^{\mu\nu}\partial_{\mu}{h}\partial_{\nu}{h}-\frac{U\left(h\right)}{\Omega^{4}}\right].
\end{equation}
Note that $ \tilde{ R}=\Omega^{-2}\left[R-\frac{6\square\Omega}{\Omega}\right] $ where $\square=\frac{1}{\sqrt{-g}}\partial_{\mu}\left(\sqrt{-g}\partial^{\mu}\right) $ \cite{Birrel:1982,Wald:1984,Synge:1955,Faraoni:1998,Lyth:t3}. This transformation (see Eqs. (\ref{E2}) and (\ref{E4})) leads to non-minimal kinetic term for the Higgs field. If we introduce a new field $\phi$, then we can get canonically normalized kinetic term by the following redefinition: \cite{Maeda:1989}
\begin{equation}
\label{E6}
\frac{d\phi}{dh}=\sqrt{\frac{\Omega^{2}+\frac{6\xi^{2}_{h}h^{2}}{M^{2}_{p}}}{\Omega^{4}}}
\end{equation}
Finally, the Lagrangian in the Einstein frame is given by
\begin{equation}
\label{E7}
\tilde{L}=\sqrt{-\tilde{g}}\left[-\frac{M^{2}_{p}}{2}\tilde{R}+\frac{1}{2}\tilde{g}^{\mu\nu}\partial_{\mu}{\phi}\partial_{\nu}{\phi}-V\left(\phi\right)\right],\hspace{.2cm}V\left(\phi\right)=\frac{U\left(h\right)}{\Omega^{4}\left(\phi\right)}.
\end{equation}
Let us discuss approximations which can be made. For small field values, $h\simeq\phi$ and $\Omega^{2}\simeq1$. Therefore, the potential for $\phi$ is the same as the initial potential for the Higgs field, i.e., $ V\left(\phi\right)=U\left(\phi\right)$. For large values of the field $ h\gg\frac{ M_{p}}{\sqrt{\xi_{h}}} $, we have the following relation:
\begin{equation}
\label{E8}
h\simeq\frac{1}{\sqrt{\xi_{h}}}\exp\left(\frac{\phi}{\sqrt{6}M_{p}}\right).
\end{equation}

Equation (\ref{E4}) implies
\begin{equation}
\label{E11}
\frac{\phi}{M_{p}}=\sqrt{\frac{3}{2}}\ln\Omega^{2},
\end{equation}
i.e.,
\begin{equation}
\label{E12}
\Omega^{2}=\exp\left(\sqrt{\frac{2}{3}}\frac{\phi}{M_{p}}\right).
\end{equation}
Since $ v\ll M_{p} $, we consider only the potential for the inflation. 
Therefore, the inflationary potential is given by following ( see Eqs.(\ref{E5}) and (\ref{E7})).
\begin{equation}
\label{E14}
V\left(\phi\right)=\frac{\lambda M^{4}_{p}}{4\xi^{2}_{h}}\left(1-\exp \left(-\sqrt{\frac{2}{3}}\frac{\phi}{M_{p}}\right)\right)^{2}
\end{equation}

It is convenient to set $ M_{p}=1 $ and consider some limits. We assumed $ \xi_{h}\gg 1 $ and $ v \xi_{h}\ll 1 $. Therefore , from Eqs. (\ref{E7}, \ref{E4} , \ref{E11})  we  estimate for large field $ \phi $. We have 
\begin{equation}
\label{E17}
\phi=\sqrt{\frac{3}{2}}\ln\left(1+\xi_{h}h^{2}\right).
\end{equation}
Then the potential $ V\left(\phi\right)=\frac{U\left(h\right)}{\left(1+\xi_{h}h^{2}\right)^{2}}=\frac{\frac{\lambda}{4}\left(h^{2}-v^{2}\right)^{2}}{\left(1+\xi_{h}h^{2}\right)^{2}} $ implies the following approximate expressions. For $ \phi<\sqrt{\frac{2}{3}}\xi^{-1}_{h}$, we have
\begin{equation}
\label{E18}
V\left(\phi\right)\approx\frac{\lambda}{4}\phi^{4}.
\end{equation}
For $\sqrt{ \frac{2}{3}}\xi^{-1}_{h}<\phi\ll\sqrt{\frac{3}{2}} $, we get
\begin{equation}
\label{E19}
V\left(\phi\right)\approx\frac{\lambda}{6\xi^{2}_{h}}\phi^{2}.
\end{equation}
Finally, as we discussed before, for $ \phi\gg\sqrt{\frac{2}{3}}\xi^{-1}_{h} $, we get Eq.(\ref{E14}).

Let us consider more complicated situation which is known as the \textit{Radiatively corrected Higgs Inflation (RHCI)}. In this case, we need to take into account corrections to $ f\left(h\right) $ and $ U\left(h\right) $ in Lagrangian (\ref{E1}). We have \cite{Barvinsky:2008,Steinwaches:2013}
\begin{equation}
\label{E20}
f\left(h\right)=\frac{M^{2}_{p}+\xi_{h}h^{2}}{2}+\frac{h^{2}}{32\pi^{2}}\textbf{C}\ln\frac{h^{2}}{\mu^{2}},U\left(h\right)=\frac{\lambda}{4}\left(h^{2}-v^{2}\right)^{2}+\frac{\lambda h^{4}}{128\pi^{2}}\textbf{A}\ln\frac{h^{2}}{\mu^{2}},
\end{equation}
where $ \Omega $ is given by
\begin{equation}
\label{E21}
\Omega^{2}\left(h\right)=1+\frac{\xi_{h}h^{2}}{M^{2}_{p}}+\frac{1}{M^{2}_{p}}\frac{h^{2}}{16\pi^{2}}\textbf{C}\ln\frac{h^{2}}{\mu^{2}}.
\end{equation}
Performing conformal transformation $ \tilde {g}_{\mu\nu}=\Omega^{2}g_{\mu\nu} $, we obtain the following relation:
\begin{equation}
\label{E22}
\frac{d\phi}{dh}=\sqrt{\frac{\Omega^{2}+\frac{6\xi^{2}_{h}h^{2}}{M^{2}_{p}}+\frac{3}{4}\frac{\xi_{h}\textbf{C}h^{2}}{\pi^{2}M^{2}_{p}}\left(1+\ln\frac{h^{2}}{\mu^{2}}\right)}{\Omega^{4}}}
\end{equation}
As a result, we find the Lagrangian density in the following form:
\begin{equation}
\label{E23}
\tilde{L}=\sqrt{-\tilde{g}}\left[-\frac{M^{2}_{p}}{2}\tilde{R}+\frac{1}{2}\left(\frac{\Omega^{2}+\frac{6\xi^{2}_{h}h^{2}}{M^{2}_{p}}+\frac{3}{4}\frac{\xi_{h}\textbf{C}h^{2}}{\pi^{2}M^{2}_{p}}\left(1+\ln\frac{h^{2}}{\mu^{2}}\right)}{\Omega^{4}}\right)\tilde{g}^{\mu\nu}\partial_{\mu}{h}\partial_{\nu}{h}-\frac{U\left(h\right)}{\Omega^{4}}\right].
\end{equation}
By using Eq. (\ref{E22}), we obtain the Lagrangian in the Einstein frame
\begin{equation}
\label{E24}
\tilde{L}=\sqrt{-\tilde{g}}\left[-\frac{M^{2}_{p}}{2}\tilde{R}+\frac{1}{2}\tilde{g}^{\mu\nu}\partial_{\mu}{\phi}\partial_{\nu}{\phi}-V\left(\phi\right)\right],\hspace{.2cm}V\left(\phi\right)=\frac{U\left(h\right)}{\Omega^{4}\left(\phi\right)}.
\end{equation}
Equations (\ref{E20}),(\ref{E21}), and (\ref{E22}) determine $ V\left(\phi\right) $ and $ \phi $
\begin{equation}
\label{E25}
V\left(\phi\right)\simeq\frac{\lambda M^{4}_{p}}{4\xi^{2}_{h}}\left[1+\frac{\textbf{A}_{I}}{16\pi^{2}}\ln\frac{h}{\mu}-\frac{2M^{2}_{p}}{\xi_{h}h^{2}}\right],
\end{equation}
where $ \textbf{A}_{I}=\textbf{A}-12\lambda $ and we use $ \textbf{C}=3 \xi_{h}\lambda $ as a one-loop correction to $ f\left(h\right) $ (see Eq.(\ref{E20})) and also we assumed $ h> \frac{M_{p}}{\sqrt{\xi_{h}}} \gg v $ and $\xi_{h}\gg 1$. Using Eqs.(\ref{E21}) and (\ref{E22}) and neglecting second order of parameter $ \textbf{C} $, we find
\begin{equation}
\label{E26}
\frac{d\phi}{dh}\simeq\frac{M_{p}\sqrt{6}\xi_{h}h}{\left(M^{2}_{p}+\xi_{h}h^{2}\right)}\left[1+\frac{\textbf{C}}{16\pi^{2}\xi_{h}}+\frac{\textbf{C}}{8\pi^{2}\xi_{h}}\frac{M^{2}_{p}}{\left(M^{2}_{p}+\xi_{h}h^{2}\right)}\ln\frac{h}{\mu}\right].
\end{equation}
Integrating the above equation, we obtain
\begin{equation}
\label{E27}
\phi=\frac{M_{p}\sqrt{6}}{2}\left[\ln\left(1+\frac{\xi_{h}h^{2}}{M^{2}_{p}}\right)+\frac{\textbf{C}}{8\pi^{2}\xi_{h}}{\frac{\frac{\xi_{h}h^{2}}{M^{2}_{p}}}{\left(1+\frac{\xi_{h}h^{2}}{M^{2}_{p}}\right)}}\ln\frac{h}{\mu}\right].
\end{equation}
Assuming $ \xi_{h}h^{2}\gg M^{2}_{p} $, we find
\begin{equation}
\label{E28}
\phi\simeq\frac{M_{p}\sqrt{6}}{2}\ln\left(\xi_{h}h^{2}\right)+\frac{\textbf{C}M_{p}\sqrt{6}}{32\pi^{2}\xi_{h}}\ln\left(\xi_{h}h^{2}\right).
\end{equation}

We find $ h\left(\phi\right) $ and substitute into Eq.(\ref{E25}) in order to express the effective potential in terms of $ \phi $. Equation (\ref{E27}) gives
\begin{equation}
\label{E30}
\frac{\xi_{h}h^{2}}{M^{2}_{p}}=\left(\exp\left(\sqrt{\frac{2}{3}}\frac{\phi}{M_{p}}\right)-1\right)\left[1-\frac{\textbf{C}}{16\pi^{2}\xi_{h}}\ln\left[\left(\exp\left(\sqrt{\frac{2}{3}}\frac{\phi}{M_{p}}\right)-1\right)/\frac{\xi_{h}\mu^{2}}{M^{2}_{p}}\right]\right].
\end{equation}
For large $\phi$, we have
\begin{equation}
\label{E31}
h\left(\phi\right)=\frac{1}{\sqrt{\xi_{h}}}\left[\exp\left(\frac{\phi}{M_{p}\sqrt{6}B}\right)\right]
\end{equation}
\begin{figure}
	\label{F1}
	\includegraphics[width=.4\textwidth]{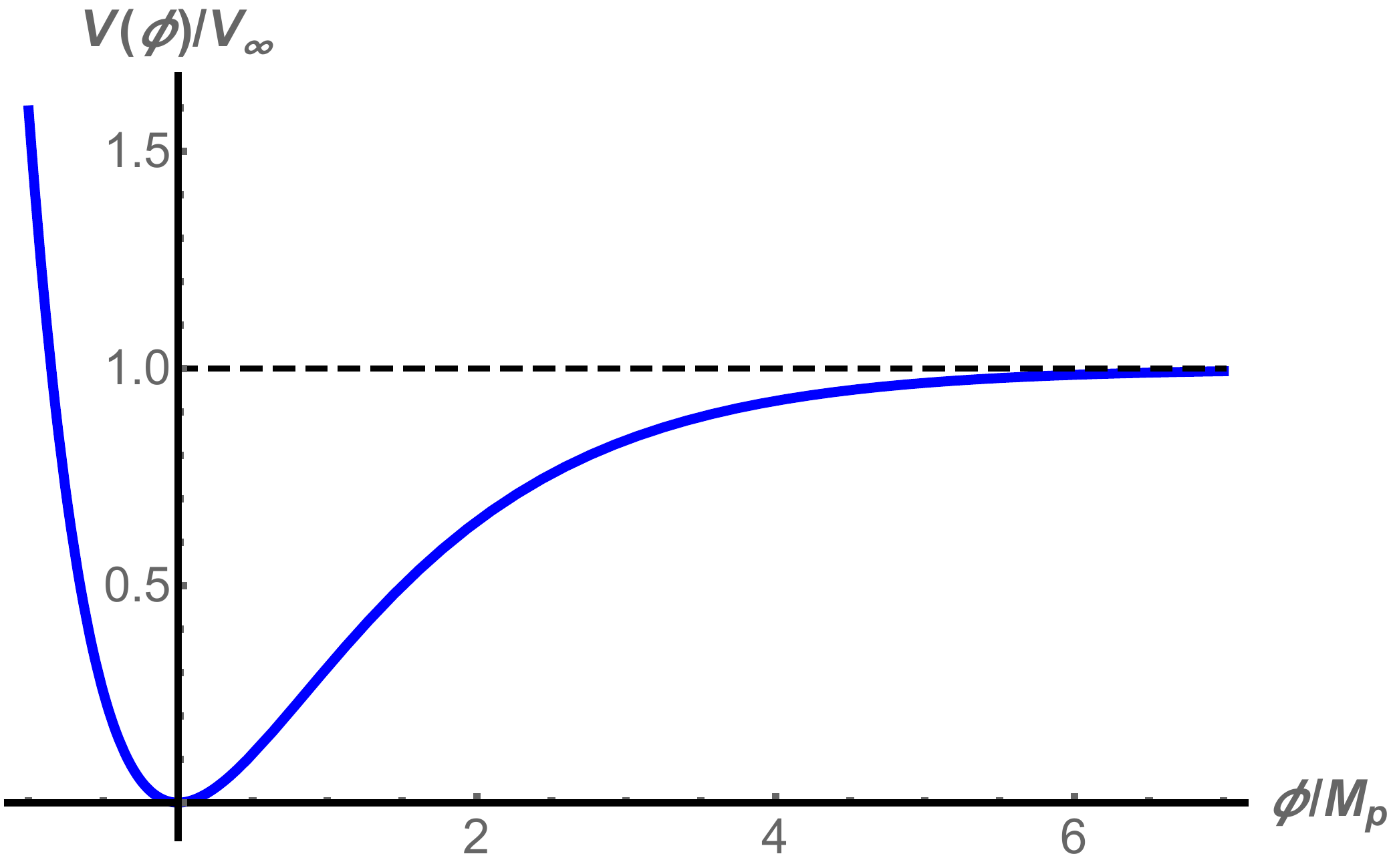}
	\includegraphics[width=.4\textwidth]{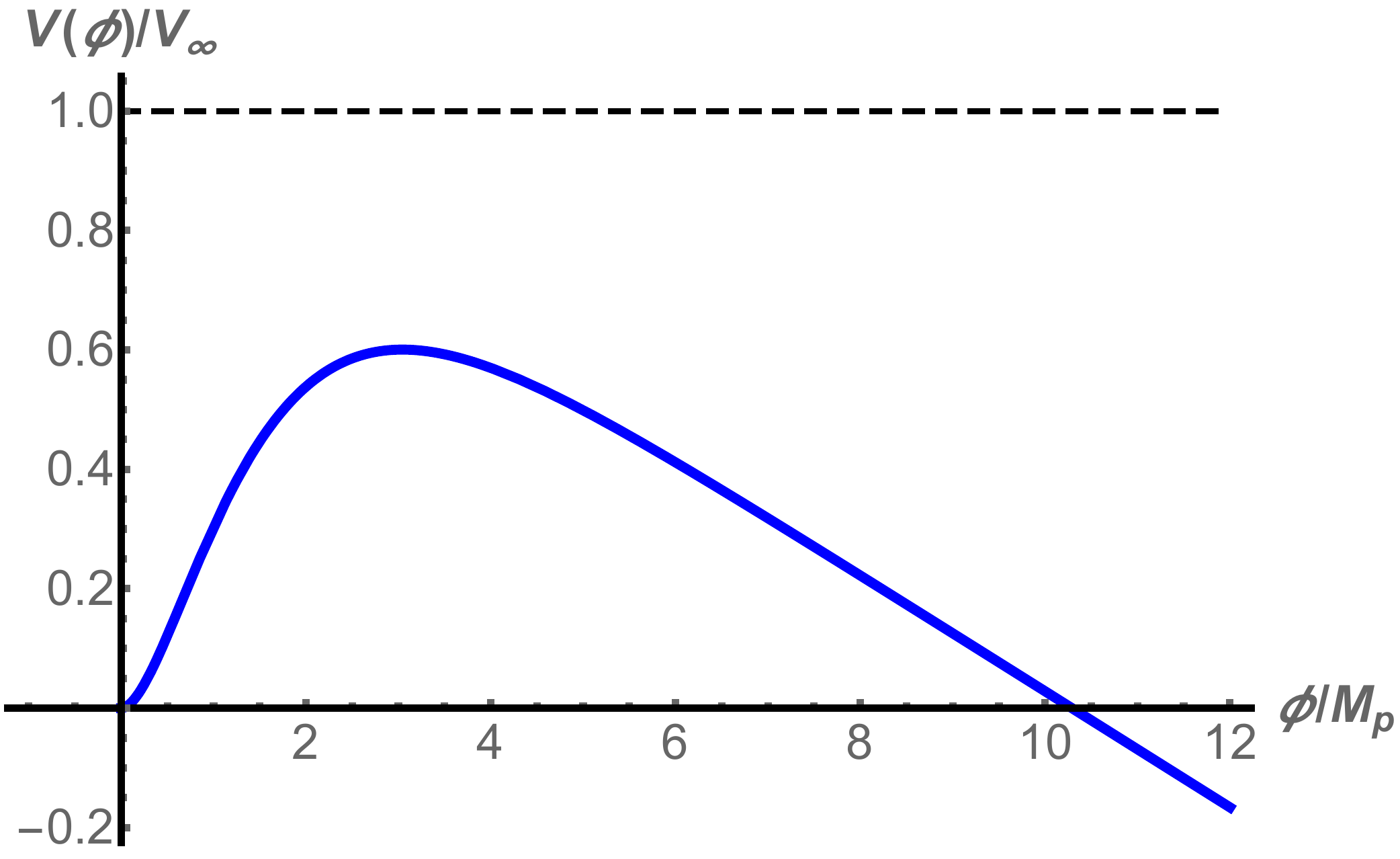}
	\caption{Left and right. Potential of Eqs. (\ref{E14}) and (\ref{E33}) for $ \mu=1.3\times10^{-5} $ and $  \textbf{A}_{I}=-37.59 $}
\end{figure}
Where $ B=1+\frac{\textbf{C}}{16\pi^{2}\xi_{h}} $ and $ \phi=M_{p}\sqrt{6}\ln\left(\sqrt{\xi_{h}}h\right)B $. This is exactly Eq.(\ref{E8}) if we neglect the effect of quantum correction $ \textbf{C} $.
 
We need an equation valid for the whole period of inflation. In order to obtain such  equation one needs to look at Eq. (\ref{E24}) as mentioned earlier  and  to obtain potential to the linear order of $\textbf{C}$. We obtain (see Eq.(\ref{E24})) the following relation for the potential:
\begin{equation}
\label{E32}
V\left(\phi\right)=\Omega^{-4}\frac{\lambda}{4}\frac{M^{4}_{p}}{\xi^{2}_{h}}\left(e^{\sqrt{\frac{2}{3}}\frac{\phi}{M_{p}}}-1\right)^{2}\left[1-\frac{\textbf{C}}{8\pi^{2}\xi_{h}}\ln\left(e^{\sqrt{\frac{2}{3}}\frac{\phi}{M_{p}}}-1\right)+\frac{\textbf{A}}{32\pi^{2}}\ln\left(e^{\sqrt{\frac{2}{3}}\frac{\phi}{M_{p}}}-1\right)\right],\Omega^{-4}=e^{-2\sqrt{\frac{2}{3}}\frac{\phi}{M_{p}}}.
\end{equation}
The above equation shows potential to  the first order in $ \textbf{C} $, where $ \Omega^{-4}=e^{-2\sqrt{\frac{2}{3}}\frac{\phi}{M_{p}}} $ and used $ \mu=\frac{M_{p}}{\sqrt{\xi_{h}}} $. To obtain the above equation, we used $ \xi_{h}\gg1 $, $ \frac{\xi^{2}_{h}h^{2}}{M^{2}_{p}}\gg 1 $, and $ v\ll h $.
We neglected also all terms with $ \textbf{C}\times\textbf{A} $ because $\textbf{A} $ is small. Using $ \Omega^{-4}=e^{-2\sqrt{\frac{2}{3}}\frac{\phi}{M_{p}}} $ and $\textbf{C}=3 \xi_{h}\lambda $, we obtain the final expression for the potential
\begin{equation}
\label{E33}
V\left(\phi\right)=\frac{\lambda}{4}\frac{M^{4}_{p}}{\xi^{2}_{h}}\left(1-e^{-\sqrt{\frac{2}{3}}\frac{\phi}{M_{p}}}\right)^{2}\left[1+\frac{\textbf{A}_{I}}{32\pi^{2}}\ln\left(e^{\sqrt{\frac{2}{3}}\frac{\phi}{M_{p}}}-1\right)\right],
\end{equation}
where $\textbf{A}_{I}=\textbf{A}-12\lambda $. We compare this potential with the Starobinsky potential (see Eq.(\ref{E14})) provided that $ 3\mu^{2}=\frac{\lambda}{\xi^{2}_{h}} $. As a one loop correction we obtain an extra term. We  estimate the numerical value of the constant $ \textbf{A}_{I} $ and plot the two potentials in order to compare them. Then we can derive predictions for the scalar perturbations, spectral index, and tensor-to-scalar ratio. The equations of motion determine the spectrum of generated electromagnetic fields. If we assume $ \frac{\xi_{h}h^{2}}{M^{2}_{p}}\gg1 $, then we find the following potential:
\begin{equation}
\label{E34}
V\left(\phi\right)\simeq\frac{\lambda}{4}\frac{M^{4}_{p}}{\xi^{2}_{h}}\left[1+\frac{\textbf{A}_{I}}{32\pi^{2}}\sqrt{\frac{2}{3}}\frac{\phi}{M_{p}}-2e^{-\sqrt{\frac{2}{3}}\frac{\phi}{M_{p}}}\right],
\end{equation}
which coincides with the equation obtained in Ref. \cite{Martin:2014}. Potential (\ref{E33}) is applicable even at the end of inflation unlike the well-known potential 
\cite{Martin:2014}.
Let us calculate the numerical value of $ \textbf{A}_{I}$ in order to plot potentials both in the Starobinsky and our model. By using Refs. \cite{Steinwaches:2013,Martin:2014}, we obtain the following relation 
\cite{Buttazzo:2013}:
\begin{equation}
\label{E35}
\textbf{A}=\frac{3}{8\lambda}\left(2g^{4}+\left(g^{2}+g{\prime}^{2}\right)^{2}-16y^{4}_{t}\right)+6\lambda+O\left(\xi_{h}^{-2}\right),\textbf{C}=3\xi_{h}\lambda+O\left(\xi_{h}^{0}\right)
\end{equation}
and then we plot potential (\ref{E33}).

We see from Figs. 1 and 2 that  Radiatively Corrected Higgs Potential has a maximum. According to the left panel of Fig. 1, the potential of the Starobinsky model has asymptotic behavior. However, if we change  the scale of $ \phi $, then the potential becomes negative. 
Note that all curves in Figs. 1 and 2 are plotted for the same value of prefactor in Eqs. (\ref{E14}), (\ref{E33}), and (\ref{E34}). In this case we obtain $\xi_{h}=1.6\times 10^{4} $, which is in accordance with our assumptions for inflation. It is useful to calculate the slow-roll parameters for potential (\ref{E33}). We have \cite{Liddle:1994}
\begin{equation}
\label{E36}
\epsilon=\frac{M^{2}_{p}}{2}\left(\frac{V_{,\phi}}{V}\right)^{2}=\frac{4}{3}\left(\frac{1+\frac{\textbf{A}_{I} e^{\sqrt{2/3}\frac{\phi}{M_{p}}}}{64\pi^{2}}}{e^{\sqrt{2/3}\frac{\phi}{M_{p}}}-1}\right)^{2}, \eta=M^{2}_{p}\frac{V_{,\phi\phi}}{V}=\frac{4}{3}\frac{\left(2-e^{\sqrt{\frac{2}{3}}\frac{\phi}{M_{p}}}\right)}{\left(e^{\sqrt{\frac{2}{3}}\frac{\phi}{M_{p}}}-1\right)^{2}}+\frac{\frac{\textbf{A}_{I}}{16\pi^{2}}e^{\sqrt{2/3}\frac{\phi}{M_{p}}}}{\left(e^{\sqrt{2/3}\frac{\phi}{M_{p}}}-1\right)^{2}}.
\end{equation} 
At the end of inflation, $ \epsilon=1 $. Therefore, for end of inflation scalar field obeys the following relation:
\begin{equation}
\label{E37}
\phi_{e}=\sqrt{\frac{3}{2}}M_{p}\ln\left[\frac{1+2/\sqrt{3}}{1-\frac{\textbf{A}_{I}}{\sqrt{3}32\pi^{2}}}\right]
\end{equation}
and the number of e-folds is given by
\begin{equation}
\label{E38}
N=-\frac{3}{4}\sqrt{\frac{2}{3}}x\mid^{x_{e}}_{x}+\frac{3}{4}\ln\left(1+\frac{\textbf{A}_{I} e^{\sqrt{2/3}x}}{32\pi^{2}}\right)\mid^{x_{e}}_{x}-\frac{3}{4}\left(e^{-\sqrt{2/3}x}\left(1+\frac{\textbf{A}_{I} e^{\sqrt{2/3}x}}{64\pi^{2}}\right)^{-1}\right)\mid^{x_{e}}_{x},
\end{equation}
where $ x=\frac{\phi}{M_{p}} $ and $ x_{e}=\frac{\phi_{e}}{M_{p}} $. In view of the shape of our potential (see Eq.(\ref{E33})) and Fig. 2, we are not able to use relations (\ref{E36})-(\ref{E38}). They are obtained in first order of $ \textbf{A}_{I} $. The requirement of slow-rolling of the inflation field imposes us to obtain more accurate relations
\begin{equation}
\label{E39}
\epsilon=\frac{4}{3}\left(\frac{1+\frac{\textbf{A}_{I} e^{\sqrt{2/3}\frac{\phi}{M_{p}}}\left[1+\frac{\textbf{A}_{I}}{32\pi^{2}}\ln\left(e^{\sqrt{2/3}\frac{\phi}{M_{p}}}-1\right)\right]^{-1}}{64\pi^{2}}}{e^{\sqrt{2/3}\frac{\phi}{M_{p}}}-1}\right)^{2}, \eta=\frac{4}{3}\frac{\left(2-e^{\sqrt{\frac{2}{3}}\frac{\phi}{M_{p}}}\right)}{\left(e^{\sqrt{\frac{2}{3}}\frac{\phi}{M_{p}}}-1\right)^{2}}+\frac{\frac{\textbf{A}_{I}}{16\pi^{2}}e^{\sqrt{2/3}\frac{\phi}{M_{p}}}\left[1+\frac{\textbf{A}_{I}}{32\pi^{2}}\ln\left(e^{\sqrt{2/3}\frac{\phi}{M_{p}}}-1\right)\right]^{-1}}{\left(e^{\sqrt{2/3}\frac{\phi}{M_{p}}}-1\right)^{2}}
\end{equation}
and the number of e-folds is given by
\begin{equation}
\label{E40}
N=-\int^{x_{e}}_{x}\frac{1}{2}\sqrt{\frac{3}{2}}\left(1-e^{-\sqrt{2/3}x}\right)e^{\sqrt{2/3}x}\left[1+\frac{\textbf{A}_{I}}{64\pi^{2}}e^{\sqrt{2/3}x}\left(1+\frac{\textbf{A}_{I}}{32\pi^{2}}\ln\left(e^{\sqrt{2/3}x}-1\right)\right)^{-1}\right]^{-1}dx,
\end{equation}
where $ x=\frac{\phi}{M_{p}}$ and $ x_{e}=\frac{\phi_{e}}{M_{p}} $. The scalar perturbations spectral index and the tensor-to-scalar ratio can be expressed in terms of slow-roll parameters as follows: \cite{Martin:2014,Gorbunov:2011}
\begin{equation}
\label{E41}
n_{s}=1-6\epsilon_{*}+2\eta_{*},
\end{equation}
\begin{equation}
\label{E42}
r=16\epsilon_{*}
\end{equation}
where the quantities with $ * $ means that the corresponding quantities are calculated when the pivot scale $ k_{*} $ crosses the horizon. One needs to solve Eqs. (\ref{E39}) and (\ref{E40}) numerically and to insert into Eqs.(\ref{E41}) and (\ref{E42}) in order to compare the obtained values with observations. Calculations show only for $ -4< \textbf{A}_{I} <10 $ the predictions for slow-roll parameters are compatible with the Planck data \cite{Planck:2018}. For instance, $ r=0.00190286 ,n_{s}=0.960691 , \textbf{A}_{I}=-3 ,N_{*}=60  $.
\begin{figure}
	\label{F6}
	\includegraphics[width=.4\textwidth]{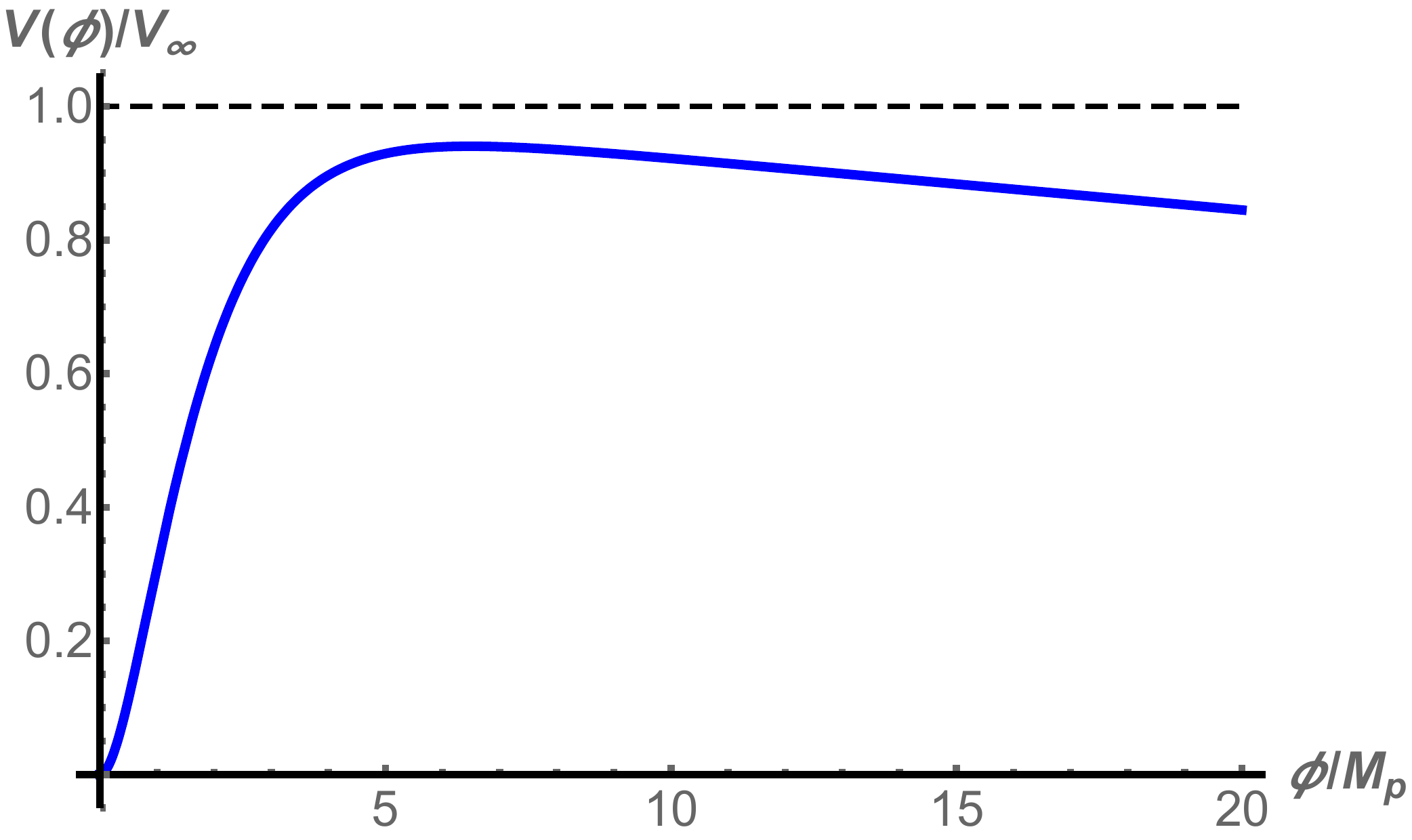}
	\includegraphics[width=.4\textwidth]{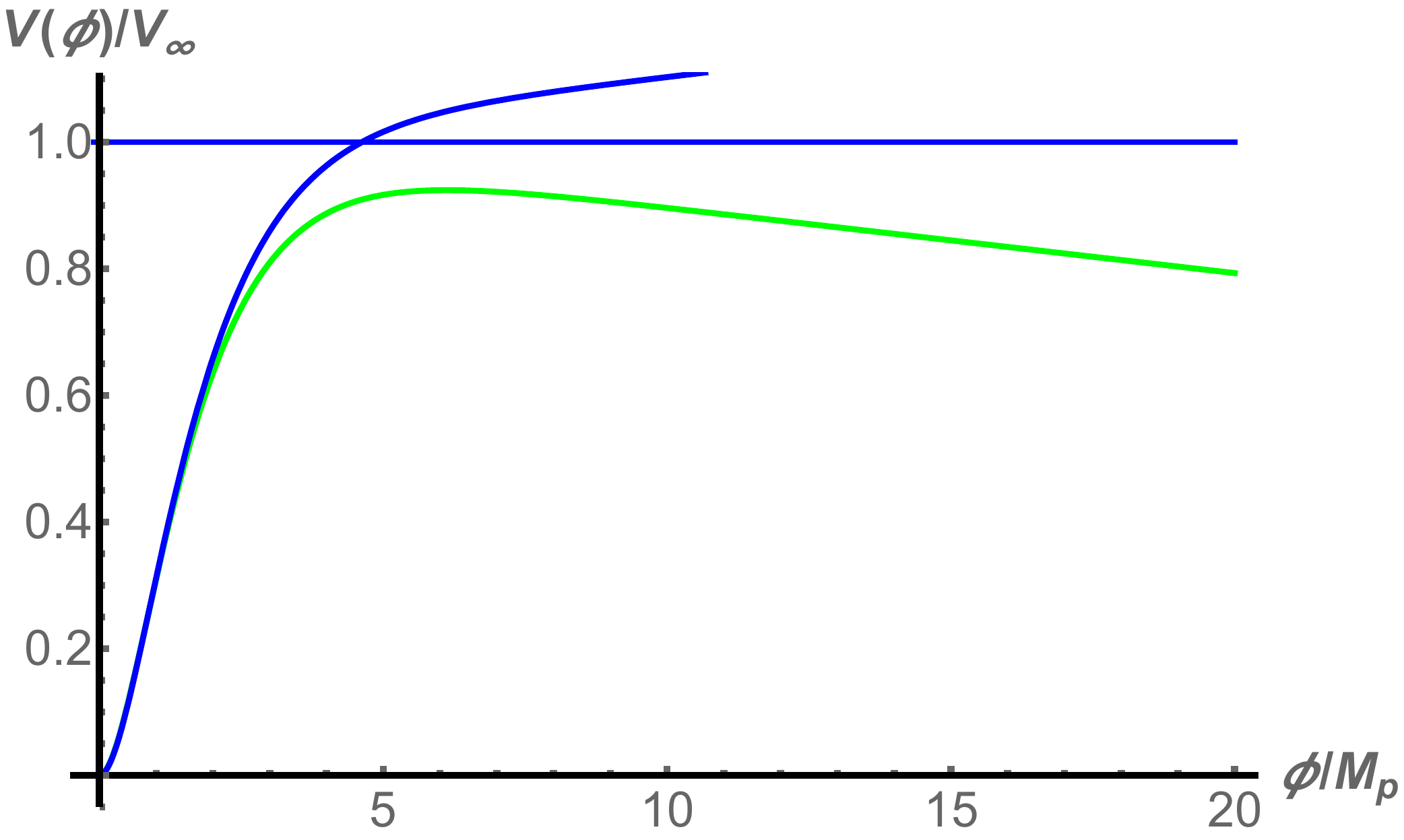}
	\caption{Left and right:Potential of Eq. ( \ref{E33}) for $ \textbf{A}_{I}=-3 $ and $ \textbf{A}_{I}=-4,+ 4, N_{*}=60 $ and $ \mu=1.3\times 10^{-5} $} 
\end{figure}

\begin{figure}[ht]
	\centering
	\includegraphics[width=0.49\textwidth]{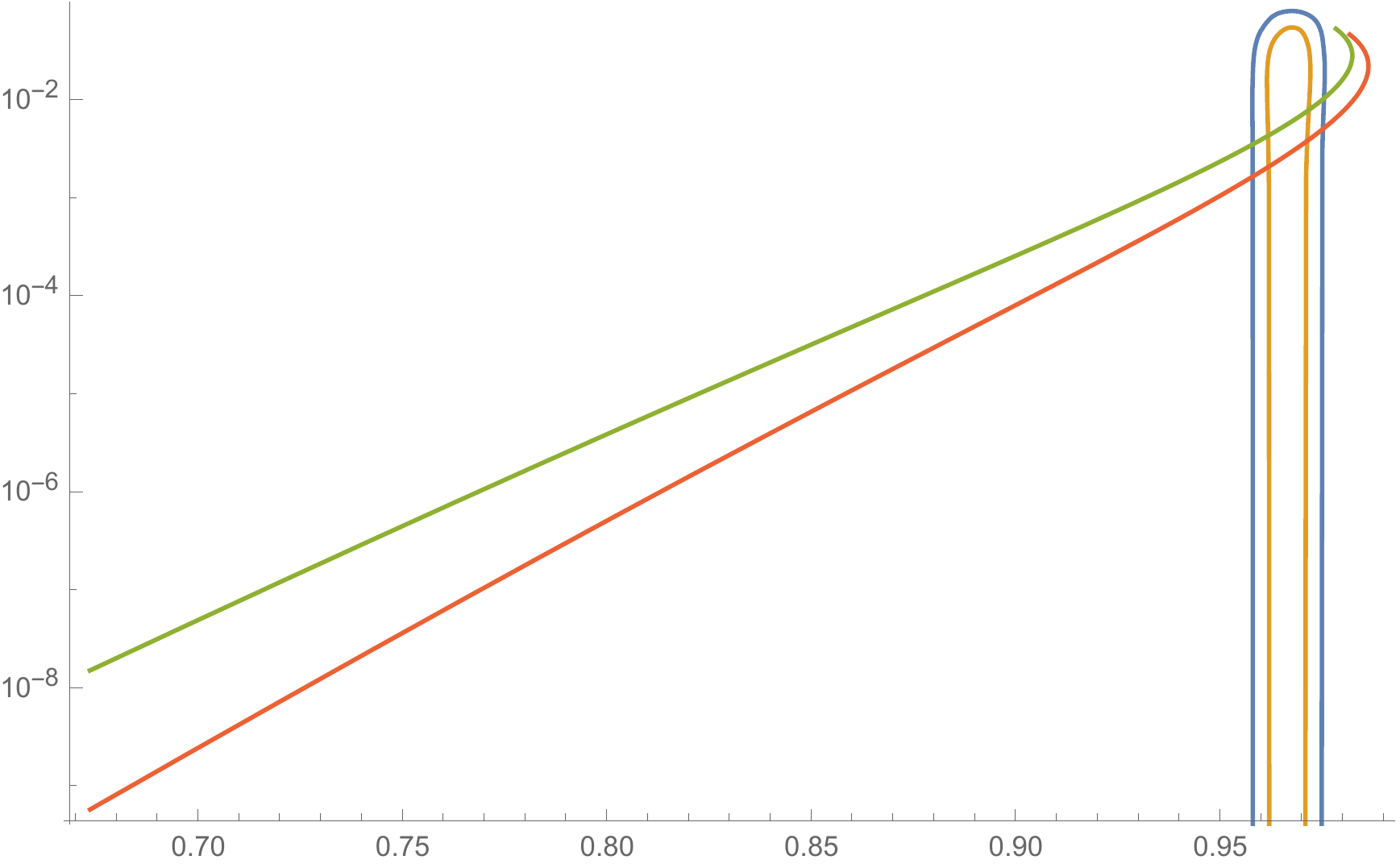}
		\caption{ Theoretical prediction for $ n_{s} $ and $  r $ in RCHI model with different values of $ -4<\textbf{A}_{I}<10 $ for $ 50<N_{*}<60 $ in comparison with Planck 2018 .  \label{F8}}
\end{figure}

We decompose $ A_{\mu} $ into its transverse and longitudinal parts. We have $ A_{\mu}=\left(A_{0},A_{i}\right)$ with $ A_{i}=A^{T}_{i}+\partial{_{i}}\chi $, where $ \partial {_{i}}A^{T}_{i}=0 $. Then the Maxwell 
action $ S=-\frac{1}{4}\int d^{4}x\sqrt{-g}F_{\mu\nu}F^{\mu\nu} $ reads as
\begin{equation}
\label{E43}
S_{0}=\frac{1}{2}\int d^{4}x\left(A^{T\prime}_{i}A^{T\prime}_{i}+A^{T}_{i}\Delta{A^{T}_{i}} \right),
\end{equation}
where $ \Delta $ is the Laplacian and $\prime $ denotes derivative with respect to the conformal time $ d\eta=\frac{dt}{a} $.

\section{Axial coupling}
\label{sec3}

As mentioned above, we consider in this paper the following axial interaction:
\begin{equation}
\label{E44}
S^{a}_{int}=-\int d^{4}x\sqrt{-g}\chi_{1}RF_{\mu\nu}\bar{F}^{\mu\nu},
\end{equation}
where $ \bar{F}^{\mu\nu}=\frac{1}{2}\epsilon^{\mu\nu\rho\sigma} F_{\rho\sigma} $ is the dual of the electromagnetic field tensor. Using conformal transformation $ \tilde{g}_{\mu\nu}=\Omega^{2}g_{\mu\nu} $, we obtain in the Einstein frame
\begin{equation}
\label{E45}
S^{a}_{int}=-\chi_{1}\int d^{4}x\sqrt{-\tilde{g}}e^{\left(\sqrt{\frac{2}{3}}\frac{\phi}{M_{p}}\right)}\left[\tilde{R}-\frac{1}{M_{p}^{2}}\left(\tilde{\nabla}\phi\right)^{2}+\frac{\sqrt{6}}{M_{p}}\tilde{\square}\phi\right]F_{\mu\nu}\bar{F}^{\mu\nu}.
\end{equation} 
If we consider only the linear approximation for electromagnetic field, then the Einstein equation gives
\begin{equation}
\label{E46}
R_{\mu\nu}-\frac{1}{2}g_{\mu\nu}R=\frac{1}{M_{p}^{2}}\left[\nabla_{\mu}\phi\nabla_{\nu}\phi-g_{\mu\nu}\left(\frac{1}{2}\left(\nabla\phi\right)^{2}-V\left(\phi\right)\right)\right].
\end{equation} 
The equation of motion for the scalar field reads
\begin{equation}
\label{E47}
\square\phi+\frac{dV\left(\phi\right)}{d\phi}=0.
\end{equation}

Using Eqs.(\ref{E46}) and (\ref{E47}), we find
\begin{equation}
\label{E49}
S^{a}_{int}=3\chi_{1}\int d^{4}x\sqrt{-\tilde{g}}e^{\left(\sqrt{\frac{2}{3}}\frac{\phi}{M_{p}}\right)}\left[\frac{1}{3M_{p}^{2}}\left(4V\left(\phi\right)\right)+\frac{\sqrt{2}}{\sqrt{3}M_{p}}\left(\frac{dV}{d\phi}\right)\right]F_{\mu\nu}\bar{F}^{\mu\nu}.
\end{equation}

In the Coulomb gauge $ A_{0}=0 , \partial_{j}A^{j}=0 $. Then Eq.(\ref{E49}) can be written as follows:
\begin{equation}
\label{E50}
S^{a}_{int}=12\chi_{1}\int d^{4}x\sqrt{-\tilde{g}}e^{\left(\sqrt{\frac{2}{3}}\frac{\phi}{M_{p}}\right)}\left[\frac{1}{3M_{p}^{2}}\left(4V\left(\phi\right)\right)+\frac{\sqrt{2}}{\sqrt{3}M_{p}}\left(\frac{dV}{d\phi}\right)\right]\epsilon_{ijk}A^{T\prime}_{i}\partial{j}A^{T}_{k},
\end{equation}
where $ F_{\mu\nu}\bar{F}^{\mu\nu}=4\epsilon_{ijk}A^{T\prime}_{i}\partial{j}A^{T}_{k} $. 

If we use Eqs.(\ref{E43}) and (\ref{E50}) so that $ S=S_{0}+S^{a}_{int} $. Then we find the equation of motion
\begin{equation}
\label{E51}
A^{T\prime\prime}_{i}-\nabla^{2}A^{T}_{i}+\omega^{\prime}\epsilon_{ijk}A^{T\prime}_{i}\partial{j}A^{T}_{k}=0,
\end{equation}
where 
\begin{equation}
\label{E52}
\omega=12\chi_{1}e^{\left(\sqrt{\frac{2}{3}}\frac{\phi}{M_{p}}\right)}\left[\frac{1}{3M_{p}^{2}}\left(4V\left(\phi\right)\right)+\frac{\sqrt{2}}{\sqrt{3}M_{p}}\left(\frac{dV}{d\phi}\right)\right].
\end{equation}
The quantization of the electromagnetic field is achieved by imposing the canonical commutation relations
\begin{equation}
\label{E53}
\left[A_{i}\left(t,\vec{x}\right),\pi_{j}\left(t,\vec{y}\right)\right]=i\int\frac{d^{3}\vec{k}}{\left(2\pi\right)^{3}}
\exp\left( i\vec{k}\cdot\left(\vec{x}-\vec{y}\right)\right)\Delta_{ij},
\end{equation}
where $\Delta_{ij}=\delta_{ij}-k_{i}k_{j}/\vec{k}^{2}$ and $\pi_{j}=\dot A_{j}$. The electromagnetic field can be expanded in terms of
the creation and annihilation operators $ a^{\dagger}_{\lambda}(\mathbf{k}) $ and $ a_{\lambda}(\mathbf{k}) $
\begin{equation}
\label{E54}
A_{i}\left(t,\vec{x}\right)=\int\frac{d^{3}k}{\left(2\pi\right)^{\frac{3}{2}}\sqrt{2k}}\sum_{\lambda=1}^{2}\epsilon_{i\lambda}
\left(\vec{k}\right)\left[a_{\lambda}\left(\vec{k}\right)\exp\left( ikx\right)+a^{\dagger}_{\lambda}
\left(\vec{k}\right)\exp\left(- ikx\right)\right].
\end{equation}
We introduce the orthogonal spatial basis as \cite{Durrer:2011}
\begin{equation}
\label{E55}
\left(\varepsilon^{\textbf{k}}_{1},\varepsilon^{\textbf{k}}_{2},\hat{\textbf{k}}\right),|\varepsilon^{\textbf{k}}_{i}|^{2}=1,\hat{\textbf{k}}=\frac{\textbf{k}}{k},\varepsilon^{\textbf{k}}_{\pm}=\frac{1}{\sqrt{2}}\left(\varepsilon^{\textbf{k}}_{1}\pm i\varepsilon^{\textbf{k}}_{2}\right)
\end{equation}
Therefore, the Fourier modes of the vector potential take the form   
\begin{equation}
\label{E56}
A^{T}_{i}\left(\eta,\textbf{k}\right)=\mathcal{A}_{+}\varepsilon_{+}+\mathcal{A}_{-}\varepsilon_{-}.
\end{equation}
We arrive at the following equation for helicity modes:
\begin{equation}
\label{E57}
\mathcal{A}^{\prime\prime}_{h}+\left[k^{2}+hk\omega^{\prime}\right]\mathcal{A}_{h}=0,
\end{equation}
where $ h=\pm$ denotes the helicity. In terms of cosmic time, Eq.(\ref{E57}) takes the form
\begin{equation}
\label{E58}
\ddot{\mathcal{A}_{h}}\left(t,k\right)+H\dot{\mathcal{A}_{h}}\left(t,k\right)+\left(\frac{k^{2}}{a^{2}\left(t\right)}+h\dot{\omega}\frac{{k}}{a}\right)\mathcal{A}_{h}\left(t,k\right)=0,
\end{equation}
where
\begin{equation}
\label{E59}
\dot{\omega}=12\chi_{1}\dot{\phi}e^{\left(\sqrt{\frac{2}{3}}\frac{\phi}{M_{p}}\right)}\left[\sqrt{\frac{2}{3}}\frac{1}{3M_{p}^{3}}\left(4V\left(\phi\right)\right)+\frac{2}{M_{p}^{2}}\left(\frac{dV}{d\phi}\right)+\frac{\sqrt{2}}{\sqrt{3}M_{p}}\frac{d^{2}V}{d\phi^{2}}\right].
\end{equation}
Taking the first and second derivative of Eq. (\ref{E33}) and substitute them into Eq. (\ref{E59}), we obtain
\begin{equation}
\label{E60}
\dot{\omega}=V_{0}\sqrt{\frac{2}{3}}\frac{1}{M^{3}_{p}}16\chi_{1}\dot{\phi}e^{\left(\sqrt{\frac{2}{3}}\frac{\phi}{M_{p}}\right)}\left[1+\frac{\textbf{A}_{I}}{32\pi^{2}}\ln\left(e^{\sqrt{\frac{2}{3}}\frac{\phi}{M_{p}}}-1\right)+\frac{3\textbf{A}_{I}}{64\pi^{2}}\right],
\end{equation}
where $ V_{0}=\frac{\lambda}{4}\frac{M_{p}^{4}}{\epsilon_{\eta}^{2}} $. All remains to be done is to numerically solve Eq. (\ref{E58}) and obtain spectrum of electromagnetic field.

\section{Power spectra of electric and magnetic field}
\label{sec4}

The power spectrum of magnetic fields is defined by
\begin{equation}
\label{E61}
\frac{d\rho_{B}}{d\ln k}=\frac{k^{3}}{\left(2\pi\right)^{2}}P_{S},
\end{equation}
where 
\begin{equation}
\label{E62}
P_{S/A}=\frac{k^{2}}{a^{4}}\left(|\mathcal A_{+}(t,k)|^{2}\pm|\mathcal A_{-}(t,k)|^{2}\right)
\end{equation}
and the upper sign corresponds to $ P_{S} $ and the lower sign to $ P_{A} $. For the electric field, the spectrum is given by
\begin{equation}
\label{E63}
\frac{d\rho_{E}}{d\ln k}=\frac{k^{3}}{\left(2\pi\right)^{2}}\frac{1}{a^{2}}\left(|\frac{\partial\mathcal A_{+}(t,k)}{\partial t}|^{2}+|\frac{\partial\mathcal A_{-}(t,k)}{\partial t}|^{2}\right).
\end{equation}
In the case of maximally helical magnetic field, $|\mathcal A_{+}|=|\mathcal A|$ with $|\mathcal A_{-}|=0 $, we have
\begin{equation}
\label{E64}
P_{S}=P_{A}=\frac{k^{2}}{a^{4}}|\mathcal A|^{2}.
\end{equation}
\begin{figure}[ht]
	\centering
	\includegraphics[width=0.53\textwidth]{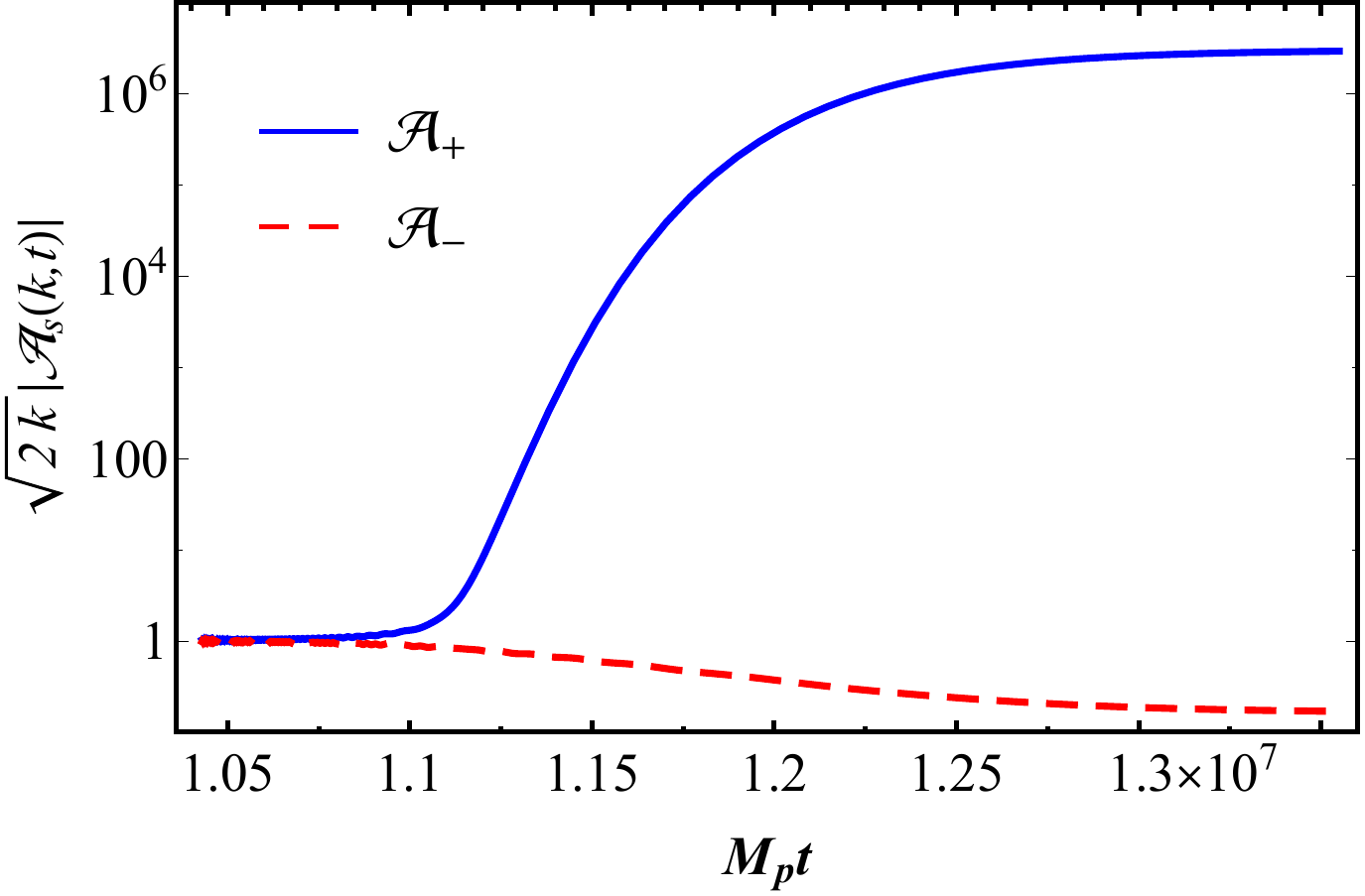}
	\caption{The time dependence of the modulus of the mode function $|\mathcal{A}_{s}(k,t)|$
		multiplied by $\sqrt{2k}$ for modes with momentum $k=10^{18}M_{p}$ and positive (blue solid line)
		or negative (red dashed line) helicity.}
	\label{fig-mode-function}
\end{figure}
\begin{figure}[ht]
	\centering
	\includegraphics[width=0.47\textwidth]{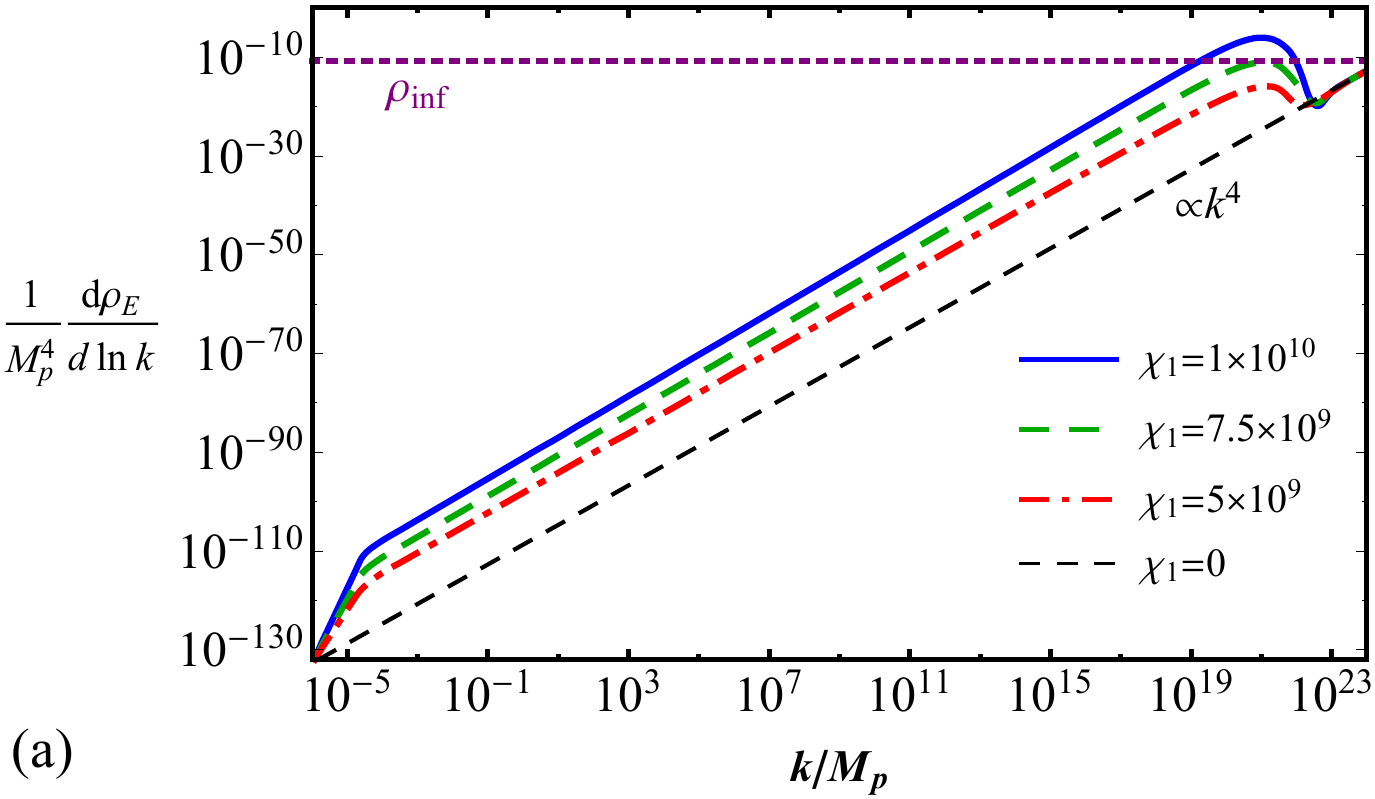}
	\hspace*{2mm}
	\includegraphics[width=0.47\textwidth]{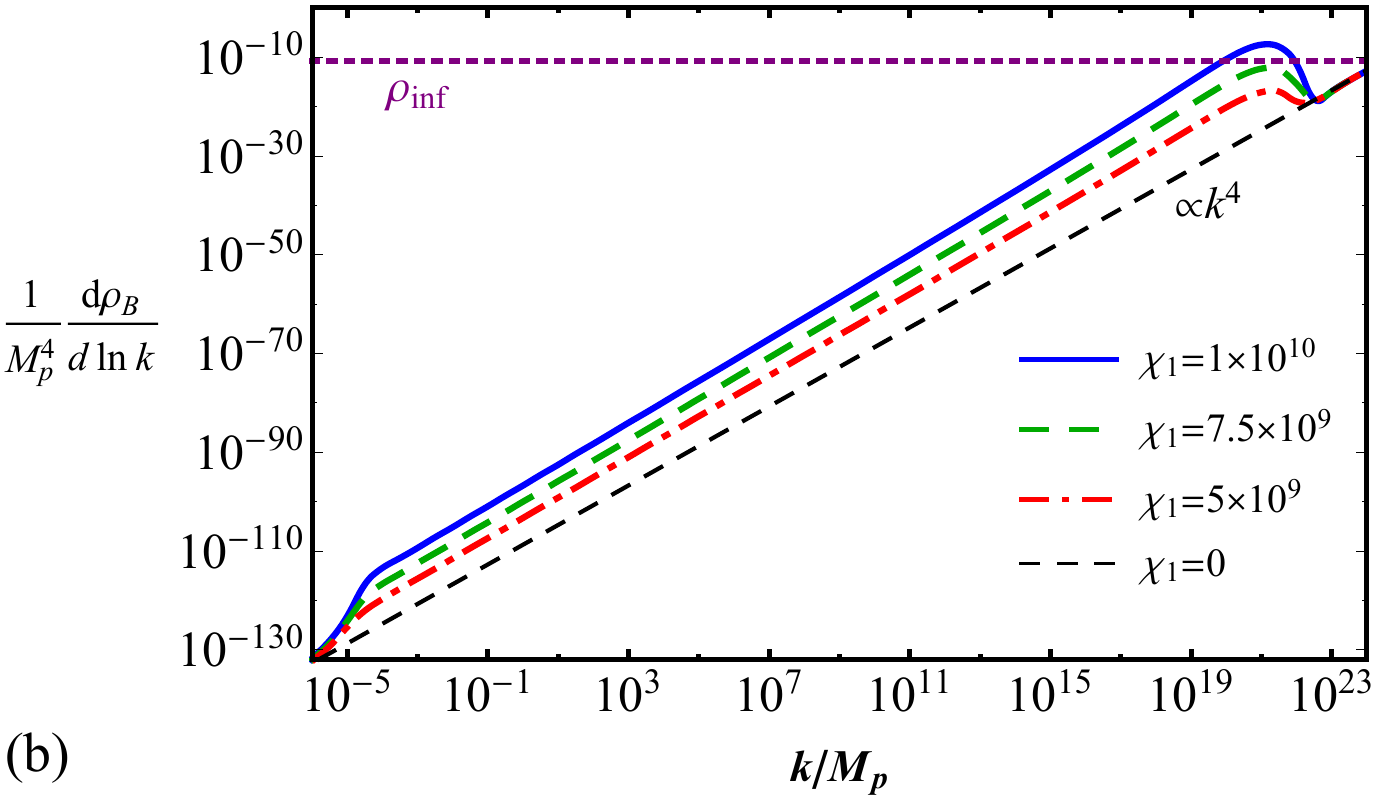}
	\caption{The spectral densities of the electric (a) and magnetic (b) energy densities 
		at the end of inflation stage as functions of the momentum for three values of the
		coupling parameter: $\chi_{1}=5\times 10^{9}$ (red dashed-dotted lines), 
		$\chi_{1}=7.5\times 10^{9}$ (green dashed lines), and $\chi_{1}=1\times 10^{10}$ (blue solid lines).
		The spectral densities of ordinary vacuum fluctuations (without amplification) are shown
		by the thin black dashed lines. The energy density of the inflation is shown by the purple
		dotted line.}
	\label{fig-spectra}
\end{figure}
\begin{figure}[ht]
	\centering
	\includegraphics[width=0.47\textwidth]{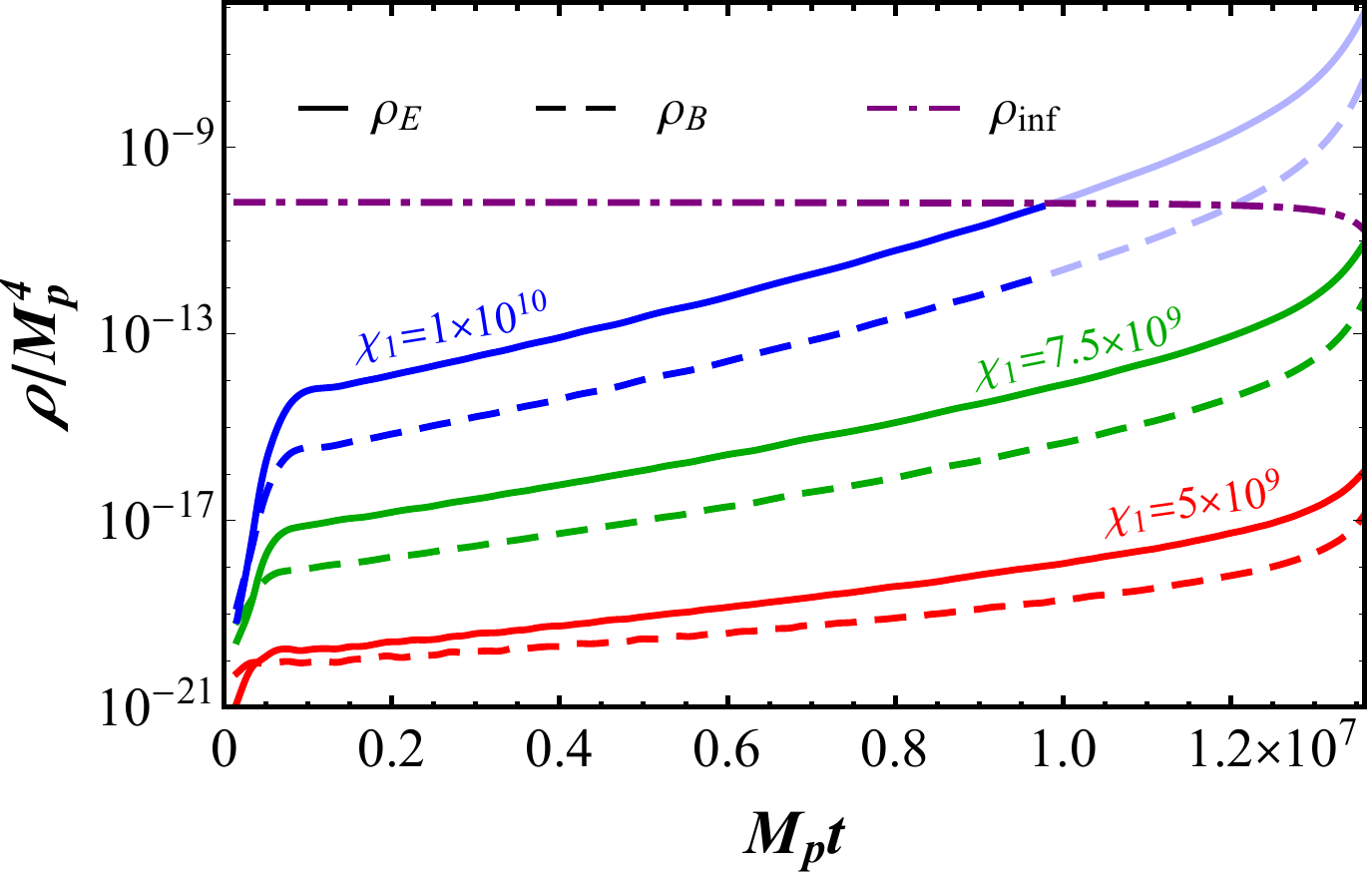}
	\caption{The time dependences of the electric (solid lines) and magnetic (dashed lines) energy densities 
		during the inflation stage for three values of the
		coupling parameter: $\chi_{1}=5\times 10^{9}$ (red lines), 
		$\chi_{1}=7.5\times 10^{9}$ (green lines), and $\chi_{1}=1\times 10^{10}$ (blue lines).
		The curves for $\chi_{1}=1\times 10^{10}$ in the back-reaction regime are shown in pale blue color.
		The energy density of the inflation is shown by the purple
		dashed-dotted line.}
	\label{fig-energy-densities}
\end{figure}

\section{Numerical calculations}
\label{sec5}

We set $ A_{I}=-3 $,$ V_{0}=7.04914*10^{-11} $, $ \phi_{0}=5.20568$, and $ N=60 $. For relevant values of $\chi$ and $k$ we solve Eq. (\ref{E58}) in order to plot the spectrum of magnetic field (see Eq.(\ref{E61})).

Figure~\ref{fig-mode-function} shows how the modes with certain value of momentum 
$k=10^{18}M_{p}$ and different helicity evolve in time.
Before the horizon crossing both polarizations oscillate in time with constant amplitude
representing the Bunch-Davies mode function. In this regime, $\sqrt{2k}|\mathcal{A}_{s}(t,k)|\approx 1$.
The situation changes drastically after the mode exits the horizon.
The mode of one polarization ($\mathcal{A}_{+}$ in our case, see the blue solid line in Fig.~\ref{fig-mode-function})
undergoes amplification, while the other one ($\mathcal{A}_{-}$, see the red dashed line in Fig.~\ref{fig-mode-function})
diminishes. This is a consequence of the axion-like coupling of the EM field to the inflation. 
As a result, an electromagnetic field with nontrivial helicity is generated.

The power spectra of the generated electric and magnetic fields at the end of inflation are
shown in Fig.~\ref{fig-spectra} for three values of the coupling parameter $\chi_{1}$.
According to this figure, we conclude that the spectra are blue with the spectral index very close to the 
unperturbed value $n_{B}\simeq 4$. The amplitude of the spectrum strongly depends on the value
of the coupling parameter. For example, increasing $\chi_{1}$ only by 2 times, we obtain 10 oreders of magnitude
of amplification.
It is important to note that for large values of the coupling parameter (e.g., for $\chi_{1}=10^{10}$)
the energy density of the electromagnetic field exceeds that of the inflation. This means that the 
back-reaction of electromagnetic fields on the background evolution cannot be neglected
and has to be self-consistently taken into account. 
For two other values, $\chi_{1}=5\times 10^{9}$ and $\chi_{1}=7.5\times 10^{9}$, the back-reaction is weak
and our analysis is valid.
The maximum in the spectrum is observed for the mode $k_{max}\sim 10^{21} M_{p}$. This corresponds to
the following correlation length of the magnetic field at the end of inflation
\begin{equation}
\lambda_{B}(t_{e})=\frac{2\pi a(t_{e})}{k_{max}}\sim 10^{6}\,M_{p}^{-1},
\end{equation}
where $a(t_{e})\approx e^{60}$ is the scale factor at the end of inflation.

Finally, integrating the spectral densities over the range of modes which exit the horizon from the
beginning of inflation until a given moment of time, we calculate the total electric and magnetic
energy densities. Their time dependences are shown in Fig.~\ref{fig-energy-densities} for three
values of the coupling parameter: $\chi_{1}=5\times 10^{9}$ (red lines), 
$\chi_{1}=7.5\times 10^{9}$ (green lines), and $\chi_{1}=1\times 10^{10}$ (blue lines). 
First, we see that the electric energy density is always greater than the magnetic one. 
Therefore, strong electric fields are generated during inflation together with magnetic ones and
the Schwinger pair production may be important for magnetogenesis. This issue, however, deserves a
separate investigation and has to be addressed elsewhere.
At second, our approach, which does not take into account the back-reaction of generated fields,
is applicable only when the electromagnetic energy density is much less than that of the inflation
(shown in Fig.~\ref{fig-energy-densities} by the purple dashed-dotted line). For $\chi_{1}=1\times 10^{10}$, and
e.g., at $t\approx 9.7\times 10^{6} M_{p}^{-1}$ the electric energy density becomes equal to that of the inflation
and the back-reaction regime occurs. The curves after this moment of time are shown in pale blue color
and do not describe the correct time dependences any more. 
Previous studies of the back-reaction regime showed that at this moment of time
the generation of electric and magnetic fields should stop and the energy densities should remain almost constant until the end 
of inflation.
Thus, the maximal possible energy density of the magnetic field generated during inflation can be
estimated by the value at the time when the back-reaction occurs, $\rho_{B,\,max}\approx 5\times 10^{-12} M_{p}^{4}$.

The important question is the post-inflationary evolution of the generated electromagnetic fields.
The electric field quickly dissipates in the highly conducting medium produced during reheating.
However, the magnetic fields can survive until the present time. Moreover, due to the nontrivial helicity
they undergo the inverse cascade process in the turbulent plasma which can strongly increase their
correlation length. 
It was shown in Ref.~\cite{SGV:2019} that the maximally helical magnetic fields with the energy density
$\rho_{B}\sim 10^{-12} M_{p}^{4}$ and the correlation length $\lambda_{B}\sim 10^{-6} M_{p}^{-1}$
at the end of inflation are transformed into the present day magnetic fields with the strength
$B_{0}\sim 10^{-15}$~G and the correlation length $\lambda_{B,0}\sim 1$~pc.

\section{Conclusions}
\label{sec-concl}

In this work we studied the generation of a magnetic field in radiatively Higgs inflation model. We used one loop quantum correction to the potential in Ref. \cite{Martin:2014}. The latter well known 
potential is not applicable at the end of inflation whereas our potential is applicable. We used the axial coupling in order to break the conformal invariance of the Maxwell action and produce a strong magnetic
field. We estimated the values $ -4< A_{I} < 10 $ for which our results are compatible with the Planck data \cite{Planck:2018}. We numerically solved Eq.(\ref{E58}).

Due to the axial coupling of the electromagnetic field with the inflation the mode of one polarization ($\mathcal{A}_{+}$ in our case, see the blue solid line in Fig.~\ref{fig-mode-function})
undergoes amplification, while the other one ($\mathcal{A}_{-}$, see the red dashed line in Fig.~\ref{fig-mode-function})
diminishes. Therefore, the electromagnetic field with nontrivial helicity is generated.

We found that the power spectra of the generated electric and magnetic fields at the end of inflation are blue with the spectral index very close to the 
unperturbed value $n_{B}\simeq 4$. However, the amplitude of power spectra strongly depends on the value of coupling parameter $ \chi_{1} $. For large values of $ \chi_{1} $, we found that the back-reaction occurs. 
However, for two  values, $\chi_{1}=5\times 10^{9}$ and $\chi_{1}=7.5\times 10^{9}$, the back-reaction is weak and our analysis is valid. 

In addition, we found that, first of all, the electric energy density is always greater than the magnetic one. Therefore, strong electric fields are generated during inflation together with magnetic ones and the 
Schwinger pair production may be important for magnetogenesis. Second, since we avoided the back-reaction problem, our calculations and the method are applicable only when the electromagnetic energy density is 
much less than that of the inflation. At $t\approx 9.7\times 10^{6} M_{p}^{-1}$ the electric energy density becomes equal to that of the inflation
and the back-reaction regime occurs.
 
We found that the maximal possible energy density of magnetic field generated during inflation can be estimated as $ \rho_{B,\,max}\approx 5\times 10^{-12} M_{p}^{4} $. This estimate is done at the time when the 
back-reaction occurs.

\begin{acknowledgments}
The author is thankful to S. Vilchinskii, E.V. Gorbar, and O. Sobol for critical comments and useful discussions during the preparation of manuscript. The author is also thankful to O.Sobol for his assistance in plotting 
figures.

\end{acknowledgments}


\end{document}